\documentclass[conference]{IEEEtran}
\IEEEoverridecommandlockouts
\usepackage{amsmath,amssymb,amsfonts}
\usepackage{graphicx}
\usepackage{textcomp}
\usepackage{booktabs}
\usepackage{makecell}
\usepackage{multirow}
\usepackage{url}
\usepackage{cite}
\usepackage{xcolor}
\usepackage{CJKutf8}
\usepackage[CJKbookmarks=true]{hyperref}

\renewcommand{\textbf}[1]{{\fontfamily{ptm}\fontseries{b}\selectfont #1}}
\renewcommand{\textit}[1]{{\fontfamily{ptm}\fontshape{it}\selectfont #1}}

\def\BibTeX{{\rm B\kern-.05em{\sc i\kern-.025em b}\kern-.08em
    T\kern-.1667em\lower.7ex\hbox{E}\kern-.125emX}}

\begin{document}

\title{PS4: Proxy-Supervised Joint Training for Real Target Speaker Extraction}

\author{
  \IEEEauthorblockN{Wanyi Ning$^{1,2}$,
  Wei Zhou$^{1}$,
  Yingpeng Li$^1$,
  Yinshang Guo$^3$,
  Haitao Qian$^1$,
  Yiming Cheng$^1$}
  \IEEEauthorblockA{\textit{$^1$Yijiahe AI, Nanjing, China \quad
  $^2$Tianjin University, Tianjin, China \quad
  $^3$Nanjing University, Nanjing, China} \\
  ningwanyi@126.com
  }
}

\maketitle

\begin{abstract}
Training target speaker extraction (TSE) models for real conversational mixtures remains challenging because large-scale training corpora and clean target speech for supervision are unavailable. 
We present PS4, a proxy-supervised training framework for TSE in real conversational mixtures, with two main contributions.
First, we construct a large-scale corpus of 71,771 training samples derived from four public datasets, covering both Chinese and English scenarios.
Each sample contains an overlapping speech mixture, per-speaker enrollment audio, a ground-truth transcript, and frame-level voice activity labels.
Second, we propose a proxy-supervised joint training strategy that fine-tunes a BSRNN-based TSE model using four complementary differentiable objectives: ASR cross-entropy, speaker similarity, frame-level voice activity detection, and perceptual audio quality.
Starting from a publicly available pre-trained checkpoint, only the BSRNN separator is updated during fine-tuning. On the REAL-T challenge\footnote{\url{https://real-tse.github.io/challenge/}} leaderboard, PS4 ranks 2nd overall, achieving the best speaker similarity and timing F1 among all submitted systems.
\end{abstract}


\section{Introduction}
\label{sec:intro}
Target speaker extraction (TSE) aims to isolate the speech of a specific speaker from a multi-talker mixture given a short enrollment utterance as a reference\cite{wang2018voicefilter,he2020speakerfilter,zmolikova2023neural,sato2021multimodal,liu2024improving}.
It has attracted increasing interest as a core component of personalized speech interfaces, meeting transcription, and assistive hearing systems.
State-of-the-art TSE models\cite{luo2019conv,luo2020dual,subakan2021attention,zeng2025usef,ling2026mc} have achieved impressive results on standard benchmarks such as VoxCeleb\cite{nagrani2017voxceleb}, WSJ0-2mix\cite{hershey2016deep} and LibriMix\cite{cosentino2020librimix}. However, these benchmarks are constructed by artificially mixing clean single-speaker recordings.
In contrast, real conversational recordings exhibit substantially different characteristics: reverberation, background noise, device-specific distortions, and natural turn-taking patterns that produce irregular overlap durations and speaker ratios\cite{carletta2005ami,yu2022m2met,watanabe2020chime,fu2021aishell,van2019dipco,realt2025}.
Because clean reference signals for individual speakers are unavailable in real-world recordings, it is not straightforward to apply conventional signal-level supervision like SI-SNR loss\cite{luo2019conv,le2019sdr} to train TSE models on such data.
As a result, existing TSE systems are predominantly trained on simulated mixtures and may degrade when deployed in real conversational scenarios.

To bridge the gap between simulated benchmarks and real-world deployment, REAL-T~\cite{realt2025} was introduced as the first benchmark for evaluating TSE systems on real conversational mixtures. It provides carefully curated real multi-talker recordings with corresponding speaker enrollment utterances and evaluation annotations, enabling systematic assessment of TSE models under realistic acoustic conditions. However, REAL-T is solely an evaluation benchmark and does not provide a matched training corpus or clean target speech for supervision. Consequently, existing TSE models, including the official REAL-T baseline, still rely on simulated mixtures for training\cite{wang2018voicefilter,he2020speakerfilter,zeng2025usef,realt2025}. How to effectively leverage large-scale real conversational recordings for TSE training therefore remains an open challenge.

In this paper, we present PS4, a proxy-supervised training framework for target speaker extraction from real conversational recordings. Instead of relying on unavailable clean target speech, PS4 leverages multiple proxy supervision signals that can be obtained directly from real conversational data. Our main contributions are summarized as follows:

\begin{itemize}
    \item We construct the first large-scale proxy-supervised training corpus, REAL-PS4\footnote{\url{https://huggingface.co/datasets/TaurenMountain/REAL-PS4}}, by reformatting four public meeting and conversational speech datasets into a unified training format compatible with the REAL-T benchmark\cite{realt2025}. The resulting corpus contains 71,771 training samples covering both Chinese and English scenarios, providing overlapping speech mixtures, speaker enrollment utterances, transcripts, and voice activity annotations.

    \item We propose a proxy-supervised joint optimization framework PS4\footnote{Our code is available at \url{https://github.com/TaurenMountain/PS4}.} that enables TSE training with multiple complementary differentiable objectives from linguistic, speaker, temporal, and perceptual perspectives while remaining fully compatible with gradient optimization.

    \item The experiment results on the REAL-T benchmark\cite{realt2025} demonstrate that PS4 significantly improves the official baseline under the challenge evaluation protocol, providing an effective and practical solution for training TSE models directly on real conversational recordings.
\end{itemize}

\section{REAL-PS4 Corpus}
\label{sec:data}

\begin{figure}[t]
  \centering
  \includegraphics[width=0.98\linewidth]{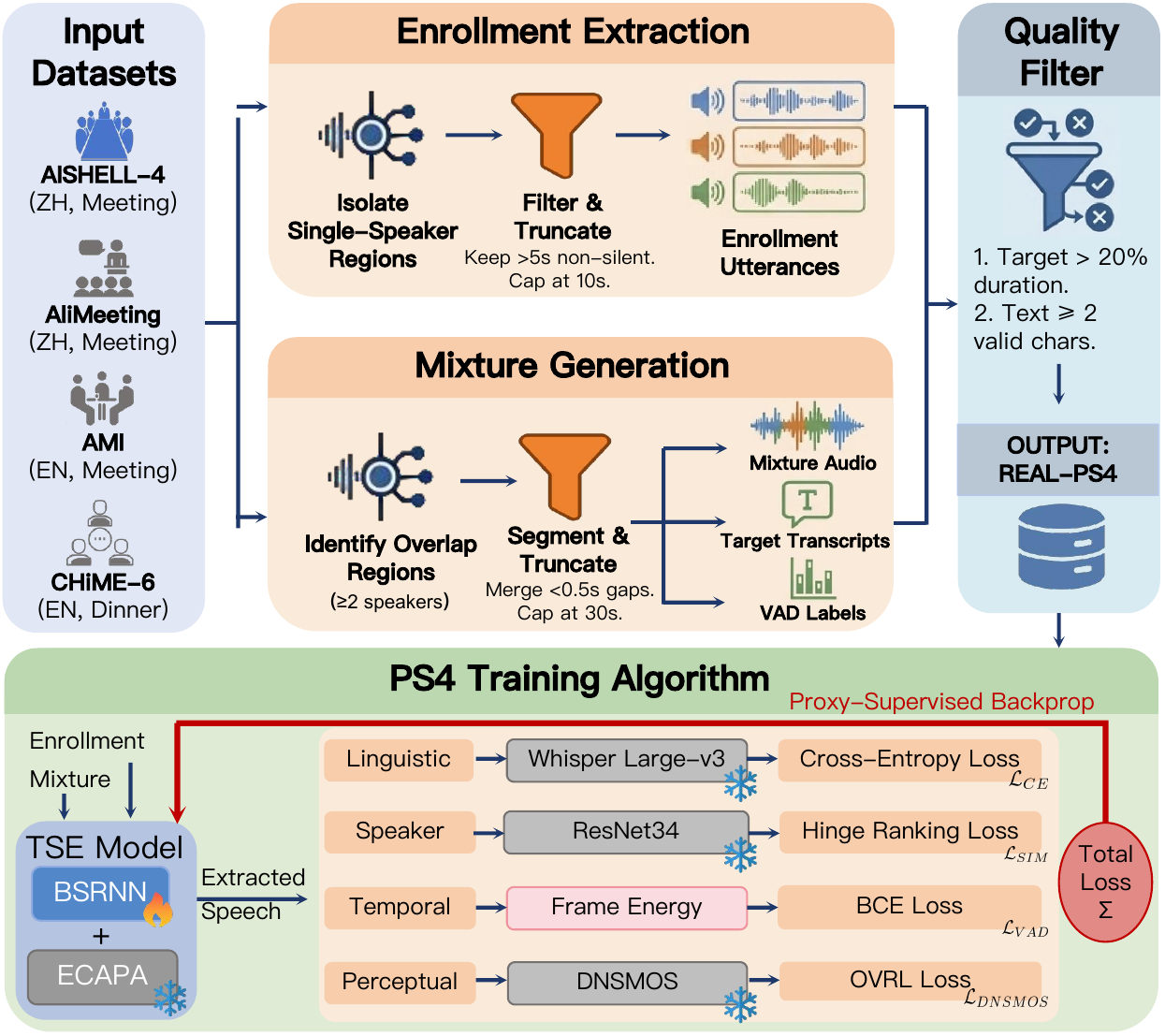}
  \caption{The workflow of PS4, including Corpus Construction and model training.}
  \label{fig:workflow}
\end{figure}

We construct a large-scale proxy-supervised training corpus, termed REAL-PS4, from four publicly available real conversational speech datasets: AISHELL-4\cite{fu2021aishell}, AliMeeting\cite{yu2022m2met}, AMI\cite{carletta2005ami}, and CHiME-6\cite{watanabe2020chime}. As shown in Fig. \ref{fig:workflow}, we generate enrollment utterances and target-speaker training samples through two parallel processing branches, followed by a final quality filtering stage.

\textbf{Enrollment extraction.}
For each recording session, speaker diarization annotations are first used to identify regions where exactly one speaker is active. These single-speaker segments are treated as candidate enrollment utterances. To ensure sufficient speaker information, only non-silent segments longer than 5\,s are retained, and up to five enrollment clips are selected for each speaker within a session.

\textbf{Mixture generation.}
Mixture samples are constructed from overlapping speech regions where two or more speakers are simultaneously active. Adjacent overlap regions separated by less than 0.5\,s are merged to avoid excessive fragmentation, while merged segments shorter than 5\,s or longer than 100\,s are discarded. For each retained segment, the corresponding mixture waveform is extracted from the original recording and peak-normalized. The target speaker's transcript is obtained from the original annotations after removing non-speech markers. In addition, frame-level voice activity detection (VAD)\cite{sohn1999statistical} labels are generated by projecting the target speaker's diarization intervals onto the extracted mixture segment, providing temporal supervision during training.


\textbf{Quality filtering.}
Finally, several quality-control rules are applied to improve the reliability of the corpus. We retain only samples in which the target speaker occupies more than 20\% of the mixture duration and the cleaned transcript contains at least two valid characters. Since Whisper\cite{radford2023whisper} accepts at most 30\,s of audio as input, mixture segments are truncated to this duration. Enrollment utterances are further limited to 10\,s to reduce GPU memory consumption during training. The remaining samples constitute the final REAL-PS4 corpus. 

\section{Method: PS4}
\label{sec:method}

We propose PS4, a proxy-supervised training framework with four complementary objectives from real conversational recordings, illustrated in Figure~\ref{fig:workflow}. Our method leverages supervision signals readily available in REAL-PS4, including transcripts, speaker enrollment, diarization annotations.
We build upon the pretrained BSRNN-ECAPA model released in the REAL-T open-source repository\footnote{\url{https://www.modelscope.cn/datasets/wenet/wesep_pretrained_models}}~\cite{realt2025}. It comprises a BSRNN separator~\cite{yu2022high} and a pretrained ECAPA-TDNN speaker encoder~\cite{desplanques2020ecapa}, initialized from a checkpoint pretrained on VoxCeleb1~\cite{nagrani2017voxceleb}. During training, we fine-tune only the separator while keeping the speaker encoder frozen.

We jointly optimize the separator using four complementary proxy supervision objectives, which constrain the extracted speech from linguistic, speaker, temporal and perceptual perspectives, respectively. The overall training objective is defined as
\begin{equation}
\mathcal{L}
=
\lambda_{\text{ce}}\mathcal{L}_{\text{CE}}
+
\lambda_{\text{sim}}\mathcal{L}_{\text{SIM}}
+
\lambda_{\text{dns}}\mathcal{L}_{\text{DNSMOS}}
+
\lambda_{\text{vad}}\mathcal{L}_{\text{VAD}},
\label{eq:loss}
\end{equation}
where $\lambda_{\text{ce}}$, $\lambda_{\text{sim}}$, $\lambda_{\text{dns}}$, and $\lambda_{\text{vad}}$ denote the corresponding loss weights.

\textbf{Linguistic supervision.}
The extracted speech should preserve the linguistic content of the target speaker. We therefore employ a frozen Whisper large-v3\cite{radford2023whisper} model as a differentiable teacher and compute the cross-entropy loss\cite{mao2023cross} between the predicted token distribution and the ground-truth transcript:
\begin{equation}
\mathcal{L}_{\text{CE}}
=
\mathrm{CrossEntropy}
\left(
\mathrm{logits}(f_{\text{Whisper}}(\hat{s})),
y_{\text{text}}
\right),
\end{equation}
where $\hat{s}$ denotes the extracted waveform and $y_{\text{text}}$ is the reference transcript. This objective encourages the separator to produce speech that remains accurately recognizable while allowing gradients to propagate through the differentiable Whisper frontend.

\textbf{Speaker supervision.}
To preserve the target speaker identity while suppressing interfering speakers, we introduce a speaker similarity objective based on cosine-margin ranking. Specifically, the similarity between the extracted speech and the enrollment utterance is encouraged to exceed that between the original mixture and the enrollment by a predefined margin:
\begin{equation}
\mathcal{L}_{\text{SIM}}
=
\mathbb{E}
\left[
\max
\left(
0,
m-
(\cos(\hat{s},e)-\cos(x,e))
\right)
\right],
\end{equation}
where $e$ denotes the enrollment utterance, $x$ is the input mixture, and $m$ is the margin. Speaker embeddings are extracted using frozen ResNet34\footnote{\url{https://modelscope.cn/datasets/wenet/wespeaker_pretrained_models}} speaker encoders\cite{wang2023wespeaker}

\textbf{Temporal supervision.}
The diarization annotations used during corpus construction naturally provide frame-level target speaker activity labels. We exploit this information as an auxiliary temporal supervision by minimizing the binary cross-entropy between the predicted frame-level speech activity of the extracted waveform and the target VAD labels:
\begin{equation}
\mathcal{L}_{\text{VAD}}
=
\mathrm{BCE}
\left(
\sigma(\mathrm{Energy}(\hat{s})),
\mathbf{y}_{\text{VAD}}
\right),
\end{equation}
where $\mathbf{y}_{\text{VAD}}$ denotes the target speaker activity labels derived from the diarization annotations. This objective encourages the separator to produce speech that is temporally consistent with the target speaker activity.

\textbf{Perceptual supervision.}
Besides preserving linguistic and speaker information, the extracted speech should also exhibit good perceptual quality. We therefore employ a differentiable implementation of DNSMOS~\cite{reddy2021dnsmos} as an auxiliary objective,
\begin{equation}
\mathcal{L}_{\text{DNSMOS}}
=
-\mathrm{DNSMOS}_{\text{OVRL}}(\hat{s}),
\end{equation}
where $\mathrm{DNSMOS}_{\text{OVRL}}$ denotes the overall perceptual quality score predicted by the DNSMOS model\footnote{\url{https://github.com/microsoft/dns-challenge}}. This objective directly encourages higher perceptual speech quality without requiring clean reference signals.


\section{Experiments}
\label{sec:exp}

We conduct experiments on the REAL-T\cite{realt2025} development set, which contains 1,991 samples from five conversational speech corpora: AISHELL-4\cite{fu2021aishell}, AMI\cite{carletta2005ami}, AliMeeting\cite{yu2022m2met}, CHiME-6\cite{watanabe2020chime}, and DipCo\cite{van2019dipco}. We follow the official REAL-T evaluation protocol and report four metrics: token error rate (TER), speaker similarity (SIM), DNSMOS OVRL perceptual quality, and timing F1\cite{realt2025}. TER is minimized, while the remaining three metrics are maximized.
We compare PS4 with the two BSRNN-based baseline systems released by the REAL-T challenge, namely \texttt{BSRNN\_TFMAP} and \texttt{BSRNN\_EMB}. Both models are trained on simulated mixtures and have not been adapted to real conversational recordings. In addition to the development set, we also submit our system to the official REAL-T leaderboard. Note that the challenge organizers updated the perceptual quality metric from DNSMOS OVRL to DNSMOS-P808\cite{naderi2020open} for the final leaderboard evaluation. Therefore, we report DNSMOS-P808 scores for the leaderboard results.

\begin{figure}[t]
  \centering
  \includegraphics[width=0.48\textwidth]{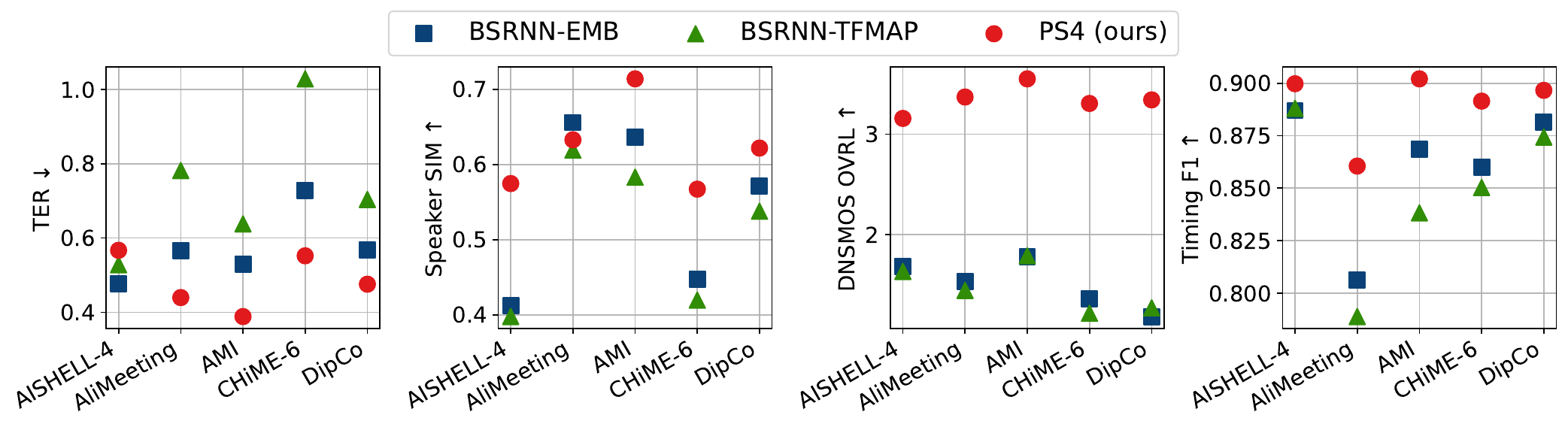}
  \caption{Per-dataset evaluation on the REAL-T development set across four metrics.}
  \label{fig:per_dataset}
\end{figure}

\begin{table}[t]
  \centering
  \caption{Per-dataset results of PS4 on the REAL-T development set.}
  \label{tab:devset}
  \setlength{\tabcolsep}{4pt}
  \begin{tabular}{lrcccc}
    \toprule
    \textbf{Dataset} & \textbf{N} & \textbf{TER}$\downarrow$ & \textbf{SIM}$\uparrow$ & \textbf{DNSMOS OVRL}$\uparrow$ & \textbf{F1}$\uparrow$ \\
    \midrule
    AISHELL-4  & 240 & 0.567 & 0.575 & 3.156 & 0.900 \\
    AliMeeting & 481 & 0.440 & 0.633 & 3.369 & 0.861 \\
    AMI        & 592 & 0.388 & 0.714 & 3.549 & 0.902 \\
    CHiME-6    & 545 & 0.552 & 0.567 & 3.305 & 0.891 \\
    DipCo      & 133 & 0.476 & 0.622 & 3.340 & 0.897 \\
    \midrule
    \textbf{Overall} & 1991 & \textbf{0.473} & \textbf{0.631} & \textbf{3.377} & \textbf{0.888} \\
    \bottomrule
  \end{tabular}
\end{table}

Fig.~\ref{fig:per_dataset} visualizes the per-dataset performance across all four metrics.
PS4 consistently outperforms both baselines on every sub-corpus, with the most pronounced gains on DNSMOS OVRL, where the baseline systems score below 2.0 while PS4 exceeds 3.1 on all five datasets.
The TER improvements are also consistent, with PS4 achieving lower error rates than both baselines across all sub-corpora.
Table~\ref{tab:devset} reports the detailed per-dataset numerical results of PS4.
AMI achieves the lowest TER of 0.388 and the highest SIM of 0.714, likely because AMI recordings have relatively clean lapel microphone channels that benefit enrollment extraction.
AISHELL-4 shows the highest TER of 0.567, which we attribute to the more challenging far-field acoustic conditions and the higher proportion of overlapping speech in that corpus.
Across all five datasets, PS4 achieves consistent SIM improvements over the mixture baseline, confirming that the speaker similarity ranking loss effectively guides the TSE model to preserve target speaker identity.

\begin{table}[t]
  \centering
  \caption{Results on REAL-T challenge leaderboard (validation set).}
  \label{tab:leaderboard}
  \begin{tabular}{lcccc}
    \toprule
    \textbf{System} & \textbf{TER}$\downarrow$ & \textbf{F1}$\uparrow$ & \textbf{SIM}$\uparrow$ & \textbf{DNSMOS-P808}$\uparrow$ \\
    \midrule
    MERL's         & \textbf{0.613} & 0.861 & 0.538 & \textbf{3.371} \\
    PS4 (ours) & 0.639 & \textbf{0.871} & \textbf{0.565} & 3.128 \\
    CARTSE's       & 0.651 & 0.857 & 0.544 & 3.138 \\
    \midrule
    BSRNN\_EMB   & 0.829 & 0.829 & 0.417 & 2.875 \\
    BSRNN\_TFMAP & 0.838 & 0.829 & 0.443 & 2.756 \\
    \bottomrule
  \end{tabular}
\end{table}

Table~\ref{tab:leaderboard} shows the challenge leaderboard results on the official validation set.
PS4 ranks 2nd overall on the leaderboard with a composite score of 3.25.
Notably, PS4 achieves the best F1 of 0.871 and the best SIM of 0.565 among all submitted systems, demonstrating superior speaker extraction accuracy and identity preservation.
Although the DNSMOS-P808 score of 3.128 is slightly lower than the top-ranked MERL's system, and TER is slightly higher, PS4 decisively outperforms MERL's on SIM and F1.
These results confirm that our four proxy supervision objectives provide effective training signals for TSE in real conversational scenarios, particularly for linguistic accuracy and speaker identity preservation.

\section{Conclusion}
\label{sec:conclusion}

In this paper, we presented PS4, a proxy-supervised joint training framework for target speaker extraction from real conversational recordings.
To address the lack of clean reference speech in real-world data, we constructed REAL-PS4, a large-scale training corpus derived from four public datasets, covering both Chinese and English scenarios.
We further proposed a joint optimization strategy that combines four complementary proxy objectives to supervise the TSE model without requiring clean isolated speech.
Experiments on the REAL-T benchmark demonstrate that PS4 consistently outperforms the official baselines across all five sub-corpora and all four evaluation metrics.
On the official challenge leaderboard, PS4 achieves the best F1 and speaker similarity scores among all submitted systems, ranking 2nd overall.
These results validate that proxy supervision provides an effective and practical alternative to conventional signal-level training for TSE in real conversational scenarios.

\bibliographystyle{IEEEtran}
\bibliography{ref}

\end{document}